 %macropackage=phyzzx
\vsize=7.5in
\hsize=6.6in
\hfuzz=20pt%this is because one equation is a bit wide.
\tolerance 10000

\baselineskip 12pt plus 1pt minus 1pt

\def\){]}
\def\({[}

\def\endlist{\par}

%\rjustline{WIS-97/XX/XXX-PH}
%\rjustline{TAUP XXXX-95}
%\rjustline{ANL-HEP-PR-97-75}

\centerline{\bf Penguins, Trees and
Final State Interactions in B Decays in Broken SU3}
\author{Harry J. Lipkin}
\smallskip
\centerline{Department of Particle Physics}
\centerline{\it Weizmann Institute of Science}
\centerline{Rehovot 76100, Israel}

%\bk
\centerline{and}

%\bk
\centerline{School of Physics and Astronomy}
\centerline{Raymond and Beverly Sackler Faculty of Exact Sciences
}
\centerline{\it Tel Aviv University}
\centerline{Tel Aviv, Israel}

\centerline{and}

\centerline{High Energy Physics Division}
\centerline{Argonne National Laboratory}
\centerline{Argonne, IL 60439-4815, USA}

\centerline{June 29, 1995}
\abstract

The availability of data on $B_s$ decays to strange quasi-two-body final
states,
either with or without charmonium opens new possibilities for understanding
different contributions of weak diagrams and in particular the relative
contributions of tree and penguin diagrams. Corresponding $B_d$ and $B_s$
decays to charge conjugate final states are equal in the SU(3) symmetry limit
and the dominant $SU(3)$ breaking mechanism is given by ratios of CKM matrix
elements. Final State Interactions effects should be small, because strong
interactions conserve $C$ and should tend to cancel in ratios between charge
conjugate states. Particularly interesting implications of decays into final
states containing $\eta$ and $\eta'$ are discussed.

\endpage

A large number of relations between ratios of $B^o$ and $B_s$ amplitudes
to charge conjugate final states are obtainable by extending the
general SU(3) symmetry relations for $B \rightarrow PP$ decays found by Gronau
et al\REF{\ROSGRO}{Michael Gronau, Jonathan L. Rosner and David London,
Phys. Rev. Lett. 73 (1994) 21; Michael Gronau, Oscar F. Hernandez, David London
and Jonathan L. Rosner, Phys. Rev. D52 (1995) 6356 and 6374}${[{\ROSGRO}]}$
beyond the two-pseudoscalar case
to all quasi-two-body charmless strange decays and charmonium strange
decays. These relations are of particular interest because: (1) they
relate a large number of decay ratios in the SU(3) symmetry limit, and
(2) strong final state interactions should cancel out in ratios between
decay amplitudes to charge conjugate final states which have the same
strong final state rescattering. We extend the treatment of ref.
${[{\ROSGRO}]}$ by noting the following points:

\pointbegin Many relations are obtainable with the U spin SU(2) subgroup
of SU(3) and in particular the discrete transformation (Weyl reflection)
$d \leftrightarrow s$ which simply interchanges the d and s flavors.
\point U spin relations can be valid also for contributions from the
electroweak penguin diagrams which break SU(3) because the photon and
the Z are both singlets under U spin (they couple equally to $d$ and $s$
quarks) while they contain octet components in SU(3) (their couplings to
$u$ quarks differs from those to $d$ and $s$).
\point Relations obtained from the discrete $d \leftrightarrow s$
transformation do not require that both final hadrons are in the same
SU(3) octet. Thus they apply equally well to other channels than PP.
\point Ratios of amplitudes that go into one another under the
$d \leftrightarrow s$ transformations and have final states which are
charge conjugates of one another should be insensitive to strong final
state interactions which are invariant under charge conjugation.
\endlist

With this approach, we find the following relations
$$ {{
A (B^o \rightarrow \pi^- K^{(*)+})
} \over {
A(B_s \rightarrow \pi^+ K^{(*)-})
}} =
{{
A (B^o \rightarrow \pi^o K^{(*)o})
} \over {
A(B_s \rightarrow \pi^o \bar K^{(*)o})
}} =
{{
A (B^o \rightarrow \rho^- K^{(*)+})
} \over {
A(B_s \rightarrow \rho^+ K^{(*)-})
}} = $$ $$ =
{{
A (B^o \rightarrow \rho^o K^{(*)o})
} \over {
A(B_s \rightarrow \rho^o \bar K^{(*)o})
}} =
{{
A (B^o \rightarrow \omega K^{(*)o})
} \over {
A(B_s \rightarrow \omega \bar K^{(*)o})
}} =
{{
A (B^o \rightarrow a_2^- K^{(*)+})
} \over {
A(B_s \rightarrow a_2^+ K^{(*)-})
}} = $$ $$ =
{{
A (B^o \rightarrow a_2^o K^{(*)o})
} \over {
A(B_s \rightarrow a_2^o \bar K^{(*)o})
}} =
{{
A (B^o \rightarrow f_2 K^{(*)o})
} \over {
A(B_s \rightarrow f_2 \bar K^{(*)o})
}} =
{{
A(B_s \rightarrow \psi \bar K^o)
} \over {
A (B^o \rightarrow \psi K^o)
}} = $$
$$ = {{
A(B_s \rightarrow \psi^{(*)} \bar K^{(*)o})
} \over {
A (B^o \rightarrow \psi^{(*)} K^{(*)o})
}} =
{{A(B_s \rightarrow D^+ \bar D_s)
} \over {
A (B^o \rightarrow  D^- D_s)
}} =
{{A(B_s \rightarrow D^{(*)+} \bar D_s^{(*)})
} \over {
A (B^o \rightarrow  D^{(*)-} D_s^{(*)})
}}  =
F_{SU3}
\eqno(1a)   $$
where $K^{(*)}$  denotes $K$ or any $K^*$ resonance,
$D^{(*)}$  denotes $D$ or any $D^*$ resonance,
$\psi^{(*)}$  denotes any charmonium state
and $F_{SU3}$ denotes an SU(3)-breaking parameter which may be different for
different final states.
Similarly for the charge conjugate states,
$$ {{
A (\bar B^o \rightarrow \pi^+ K^{(*)-})
} \over {
A(\bar B_s \rightarrow \pi^- K^{(*)+})
}} =
{{
A (\bar B^o \rightarrow \pi^o \bar K^{(*)o})
} \over {
A(\bar B_s \rightarrow \pi^o K^{(*)o})
}} =
{{
A (\bar B^o \rightarrow \rho^+ K^{(*)-})
} \over {
A(\bar B_s \rightarrow \rho^- K^{(*)+})
}} = $$ $$ =
{{
A (\bar B^o \rightarrow \rho^o \bar K^{(*)o})
} \over {
A(\bar B_s \rightarrow \rho^o K^{(*)o})
}} =
{{
A (\bar B^o \rightarrow \omega \bar K^{(*)o})
} \over {
A(\bar B_s \rightarrow \omega K^{(*)o})
}} =
{{
A (\bar B^o \rightarrow a_2^+ K^{(*)-})
} \over {
A(\bar B_s \rightarrow a_2^- K^{(*)+})
}} = $$ $$ =
{{
A (\bar B^o \rightarrow a_2^o \bar K^{(*)o})
} \over {
A(\bar B_s \rightarrow a_2^o K^{(*)o})
}} =
{{
A (\bar B^o \rightarrow f_2 \bar K^{(*)o})
} \over {
A(\bar B_s \rightarrow f_2 K^{(*)o})
}} =
{{
A(\bar B_s \rightarrow \psi K^o)
} \over {
A (\bar B^o \rightarrow \psi \bar K^o)
}} = $$
$$ = {{
A(\bar B_s \rightarrow \psi^{(*)} K^{(*)o})
} \over {
A (\bar B^o \rightarrow \psi^{(*)} \bar K^{(*)o})
}}
= {{A(\bar B_s \rightarrow D^- D_s)
} \over {
A (\bar B^o \rightarrow  D^+ \bar D_s)
}} =
{{A(\bar B_s \rightarrow D^{(*)-} D_s^{(*)})
} \over {
A (\bar B^o \rightarrow  D^{(*)+} \bar D_s^{(*)})
}} =
F_{SU3}
\eqno(1b)   $$

Note that only tree and penguin diagrams
contribute to these transitions and that the individual tree and
penguin diagrams, including both gluonic and electroweak penguins,
also go into one another under this transformation.

The final states in the numerator and denominator of each ratio go into one
another under charge conjugation. Thus final state strong
interactions which conserve $C$ should be the same and therefore not
disturb the equalities.
These ratios may then give information about the
relative contributions of different weak diagrams without the usual caveats
about unknown strong phases.

In the limit of exact SU(3) symmetry $F_{SU(3)}= 1$. Thus the relations
(1) hold when SU(3) is broken by the same factor in all cases. This is
not expected to be valid everywhere. Thus the relations (1) provide a
means for selecting groups of related decay modes which all have the same
SU(3) breaking factor.

Since experimental data generally quote branching ratios rather than partial
widths or amplitudes, we note that the relations (1b) can be rearranged to
give ratios of branching ratios from the same initial state; e.g.
$$
BR(B^o \rightarrow \pi^- K^{(*)+}) / BR(B^o \rightarrow \pi^o K^{(*)o}) /
BR(B^o \rightarrow \rho^- K^{(*)+})/ BR(B^o \rightarrow \rho^o K^{(*)o})
= $$ $$ =
BR(B_s \rightarrow \pi^+ K^{(*)-})/ / BR(B_s \rightarrow \pi^o \bar K^{(*)o})
/BR(B_s \rightarrow \rho^+ K^{(*)-}) /BR(B_s \rightarrow \rho^o \bar K^{(*)o})
\eqno(2a)   $$ $$
BR(B^o \rightarrow \omega K^{(*)o}) /BR(B^o \rightarrow a_2^- K^{(*)+})
/BR(B^o \rightarrow a_2^o K^{(*)o}) /BR(B^o \rightarrow f_2 K^{(*)o}) = $$ $$ =
BR(B_s \rightarrow \omega \bar K^{(*)o}) /BR(B_s \rightarrow a_2^+ K^{(*)-})
BR(B_s \rightarrow a_2^o \bar K^{(*)o}) /BR(B_s \rightarrow f_2 \bar K^{(*)o})
\eqno(2b)   $$
These relations can be used to distinguish between decays having the same
SU(3) breaking factor and those having different SU(3) breaking factors.

The dominant SU(3) breaking effect is in the difference between the weak
strangeness-conserving and strangeness-changing
vertices. For the charmless tree diagrams this breaking introduces
a common factor $F_{SU3} = r_{T(usd)} \equiv V_{us}/V_{ud} \approx 0.23$ into
each ratio, thereby leaving all the
ratios (1a) and (1b) equal to one another and only changing the value to
$V_{us}/V_{ud}$ instead of unity.
For the charmonium and charmed pair tree diagrams the appropriate breaking
factor $F_{SU3} = r_{T(cds)} \equiv V_{cd}/V_{cs} \approx r_{T(usd)}$
which is nearly the same as that of the charmless tree diagrams.

If only tree diagrams contribute, relations for the charmonium branching ratios
analogous to eqs. (2) can also be written.

Penguin contributions will have a different SU(3) breaking factor
$ F_{SU3} = r_P >  1  >  r_{T(usd)} \approx 0.23 $;
e.g $V_{cs}/V_{cd}$ or $V_{ts}/V_{td}$.

The penguin diagram is expected to dominate in the charmless $B^o$ decays, and
perhaps also in the charmless $B_d$ decays, since the charmless tree diagram is
Cabibbo suppressed. The tree diagram is expected to dominate in the charmonium
and charmed pair decays, where the tree is Cabibbo favored while the penguin
requires the creation of a heavy quark pair by gluons from the vacuum.
These features can be checked out by experimental tests of the relations (1).
The most interesting cases are those in which both penguin and tree
contributions are appreciable and CP violation can be observed in the
interference. These decay modes can be identified by violations of the relations
(1). The most favorable candidates seem to be the $B_s$ decays where
one of the two weak vertices is Cabibbo favored and will have a better chance
to compete with the penguin.

We can correct the relations (1) for the difference between penguin and
tree SU(3) breaking by writing for example:
$$ {{
A (B^o \rightarrow \rho^o K^{(*)o})
} \over {
A(B_s \rightarrow \rho^o \bar K^{(*)o})
}} = {{r_{T(usd)} \cdot T_s + r_P \cdot P_s}\over{T_s + P_s}};
~ ~ ~ ~ ~ ~
{{
A (\bar B^o \rightarrow \rho^o \bar K^{(*)o})
} \over {
A(\bar B_s \rightarrow \rho^o K^{(*)o})
}} = {{r_{\bar T(usd)} \cdot \bar T_s + r_{\bar P} \cdot
\bar P_s}\over{\bar T_s + \bar P_s}}
\eqno (3a)  $$
where $T$, $P$, $\bar T$ and $\bar P$  denote respectively the contributions
to the decay amplitude $A (B^o \rightarrow \rho K^{(*)o})$ and to the charge
conjugate decay amplitude $A(\bar B^o \rightarrow \rho \bar K^{(*)o})$
from tree and penguin diagrams and $T_s$, $P_s$, $\bar T_s$ and
$\bar P_s$  denote respectively the analogous contributions to the
corresponding
$B_s$ decay amplitudes. We can obtain similar relations for final states
containing the $\omega$ instead of the $\rho^o$ by noting that the
corresponding $\rho^o$ and $\omega$ decay modes are related if
electroweak penguins are neglected, because the tree diagram
produces both $\rho^o$ and $\omega$ via their common
$(u \bar u)$ component and the penguin produces both via their common
$(d \bar d)$ component\REF{\PENGRHO}{Harry J. Lipkin,
Physics Letters B353 (1995) 119}${[{\PENGRHO}]}$.
$$ {{
A (B^o \rightarrow \omega K^{(*)o})
} \over {
A(B_s \rightarrow \omega \bar K^{(*)o})
}} = {{r_{T(usd)} \cdot T_s - r_P \cdot P_s}\over{T_s - P_s}};
~ ~ ~ ~ ~ ~
{{
A (\bar B^o \rightarrow \omega \bar K^{(*)o})
} \over {
A(\bar B_s \rightarrow \omega K^{(*)o})
}} = {{r_{\bar T(usd)} \cdot \bar T_s - r_{\bar P} \cdot
\bar P_s}\over{\bar T_s - \bar P_s}}
\eqno (3b)  $$

   $$ {{BR (B^o \rightarrow K^{(*)o} \rho^o)}\over{
BR (B^o \rightarrow  K^{(*)o} \omega )}}=
\left| {{T + P}\over { T-P}}\right| ^2 ;
~ ~ ~ ~ ~ ~
   {{BR (\bar B^o \rightarrow \bar K^{(*)o} \rho^o)}\over{
BR (\bar B^o \rightarrow \bar K^{(*)o} \omega )}}=
\left| {{\bar T + \bar P}\over { \bar T-\bar P}}\right| ^2
 \eqno(4a) $$
   $$ {{BR (B_s \rightarrow \bar K^{(*)o} \rho^o)}\over{
BR (B_s \rightarrow  \bar K^{(*)o} \omega )}}=
\left| {{T_s + P_s}\over { T_s-P_s}}\right| ^2 ;
~ ~ ~ ~ ~ ~
   {{BR (\bar B_s \rightarrow K^{(*)o} \rho^o)}\over{
BR (\bar B_s \rightarrow K^{(*)o} \omega )}}=
\left| {{\bar T_s + \bar P_s}\over { \bar T_s-\bar P_s}}\right| ^2
 \eqno(4b) $$
We can also consider linear combinations that project out direct and
interference terms:
$$
{{|A (B^o \rightarrow \rho^o K^{(*)o})|^2+ |A (B^o \rightarrow
\omega K^{(*)o})|^2
} \over {
|A(B_s \rightarrow \rho^o \bar K^{(*)o})|^2+|A(B_s \rightarrow \omega
\bar K^{(*)o})|^2
}} = $$
$$ = {{|A (\bar B^o \rightarrow \rho^o \bar K^{(*)o})|^2+ |A (\bar B^o
\rightarrow \omega  \bar K^{(*)o})|^2
} \over {
|A(\bar B_s \rightarrow \rho^o K^{(*)o})|^2+|A(\bar B_s \rightarrow \omega
K^{(*)o})|^2
}} = {{|r_{T(usd)}^2 T_s^2|+ |r_P^2  P_s^2|}\over{|T_s|^2+|P_s|^2}}
\eqno (5a)  $$
$$
{{|A (B^o \rightarrow \rho^o K^{(*)o})|^2- |A (B^o \rightarrow
\omega K^{(*)o})|^2
} \over {
|A(B_s \rightarrow \rho^o \bar K^{(*)o})|^2-|A(B_s \rightarrow \omega
\bar K^{(*)o})|^2
}} = r_{T(usd)}  r_P
\eqno (5b)  $$
$$ = {{|A (\bar B^o \rightarrow \rho^o \bar K^{(*)o})|^2-
|A (\bar B^o \rightarrow \omega  \bar K^{(*)o})|^2
} \over {
|A(\bar B_s \rightarrow \rho^o K^{(*)o})|^2-|A(\bar B_s \rightarrow \omega
K^{(*)o})|^2
}} = r_{\bar T(usd)}  r_P
\eqno (5c)  $$
where we have noted that $|T_s| = |\bar T_s|$ and $|P_s| = |\bar P_s|$.

Any violation of the relations (1) could indicate
existence of both tree and penguin contributions and also offer the possibility
of measuring their relative phase. Since the penguin and tree can have
different weak phases, $CP$ violation can be observable as
asymmetries in decays of charge-conjugate $B$ mesons into
charge-conjugate final states and also in differences between the
charge-conjugate $\rho/\omega$ ratios (4a) and (4b).

The relations (4) also  provide additional
input from $B \rightarrow K \omega$ decays that can be combined with
isospin analyses of $B \rightarrow K \rho$ decays to
separate penguin and tree contributions\REF{\PBPENG}{Yosef Nir and Helen Quinn
Phys. Rev. Lett. {\bf 67 (1991)}
541; Harry J. Lipkin, Yosef Nir, Helen R. Quinn and Arthur E. Snyder,
Physical Review {\bf D44} (1991) 1454}${[{\PBPENG}]}$.
A similar additional input is obtainable from combining $\omega$
decay modes with isospin analyses of other $\rho$ decay
modes\REF{\GRONAU}{M. Gronau and D. Wyler, Phys. Lett. {\bf B265} 172 (1991);
M. Gronau and D. London, Phys. Lett. {\bf B253} 483 (1991);
I. Dunietz, Phys. Lett. {\bf B270} 74 (1991).}
${[{\GRONAU}]}$.

Similar relations, with different values of T and P hold
for the other ratios. Before extending this result to other cases, we
note that additional SU(3) breaking can arise from differences in
hadronic form factors. This can be seen at the quark level by noting
the quark couplings in the color-favored and color-suppressed tree
diagrams and penguin diagrams:
$$ B^o (b \bar d) \rightarrow (u \bar d)_{cfhad} (\bar u s)_{cfpt}
; ~ ~ ~ ~ ~ ~
B^s (b \bar s) \rightarrow (u \bar s)_{cfhad} (\bar u d)_{cfpt}
\eqno (6a) $$
$$ B^o (b \bar d) \rightarrow (c \bar d)_{cfhad} (\bar c s)_{cfpt}
; ~ ~ ~ ~ ~ ~
B^s (b \bar s) \rightarrow (c \bar s)_{cfhad} (\bar c d)_{cfpt}
\eqno (6b) $$
$$ B^o (b \bar d) \rightarrow (u \bar u)_{cspt} (\bar d s)_{cshad}
; ~ ~ ~ ~ ~ ~
B^s (b \bar s) \rightarrow (u \bar u)_{cspt} (\bar s d)_{cshad}
\eqno (7a) $$
$$ B^o (b \bar d) \rightarrow (c \bar c)_{cspt} (\bar d s)_{cshad}
; ~ ~ ~ ~ ~ ~
B^s (b \bar s) \rightarrow (c \bar c)_{cspt} (\bar s d)_{cshad}
\eqno (7b) $$
$$ B^o (b \bar d) \rightarrow_{penguin} (\bar d s) \rightarrow Hadrons
; ~ ~ ~ ~ ~ ~
B^s (b \bar s) \rightarrow_{penguin} (\bar s d) \rightarrow Hadrons
\eqno (7c) $$
where $cfpt$ and $cspt$ denote respectively color-favored and color
suppressed form factors which are point-like and proportional to wave
functions at the origin; e.g. to factors like $f_\pi$ or $f_K$, while
$cfhad$ and $cshad$
denote respectively color-favored and color suppressed form factors which
involve overlap integrals on a hadronic scale.
The pairs of color favored transitions (6) are seen to involve different form
factors. One has a hadronic nonstrange form factor and a pointlike strange
form factor; the other has a hadronic strange form factor and a pointlike
nonstrange form factor. This form factor difference has been recently pointed
out\REF{\CLOLIP12}{Frank E. Close and Harry J. Lipkin, Physics Letters B405
(1997) 157}${[{\CLOLIP12}]}$ as possibly responsible for a reversal of relative
phase of the two contributions for exclusive decay modes where there are nodes
in wave functions.

The color suppressed tree and the penguin transitions (7) are seen to involve
identical form factors in both cases. The trees have the same $u \bar u$ or
$c \bar c$ form factor and charge-conjugate hadronic $\bar s d$ and $d \bar s$
form factors. Pairs of penguin diagrams always have the same form factors,
since the hadronization into the final state occurs from charge conjugate
intermediate states of a single $q \bar q$ pair and a gluon or electroweak
boson. The only possible difference arises from the slight difference in the
hadronic scale of the $B_s$ and $B^o$ wave functions.

We therefore extend the result to all cases where form-factor
corrections are expected to be small: those having no color-favored
tree contribution and no $s \bar s$ component in the wave function as in
$\eta$ and $\eta'$.
$$
{{
A(B^o \rightarrow \pi^o  K^{*o})} \over {A (B_s \rightarrow \pi^o \bar K^{*o})
}} \approx
{{A (B^o \rightarrow \rho^o K^{*o})}\over {A(B_s \rightarrow \rho^o
\bar K^{*o})
}} \approx
{{A (B^o \rightarrow \omega K^{*o})}\over {A(B_s \rightarrow \omega
\bar K^{*o})
}} \approx
{{
A(B^o \rightarrow M^o  K^{*o})} \over {A (B_s \rightarrow M^o \bar K^{*o})
}} \approx $$
$$ \approx
{{r_{T(usd)} \cdot T_s + r_P \cdot P_s}\over{T_s + P_s}};
            \eqno(8)   $$
where $M^o$ can denote any neutral isovector or ideally mixed
nonstrange isoscalar meson; e.g. $\pi^o$, $\rho^o$, $a_2^o$ or $\omega$.
Similarly $M^{\pm}$ will denote any charged meson pair; e.g.
$\pi^{\pm}$, $\rho^{\pm}$ or $a_2^{\pm}$.
Each ratio is equal to
an expression analogous to the right hand side of (3)
with appropriate different values for $T$ and $P$.

The color-favored transitions to charged final states may have form
factor corrections. Let $F_{AB}$ denote this form factor correction,
where A and B denote the two particles in the final state. Then
$$
{{A (B^o \rightarrow \pi^- K^{*+})}\over{A(B_s \rightarrow \pi^+ K^{*-})
}} \approx
{{A (B^o \rightarrow \rho^- K^{*+})}\over{A(B_s \rightarrow \rho^+ K^{*-})
}} \approx
{{A (B^o \rightarrow M^- K^{*+})}\over{A(B_s \rightarrow M^+ K^{*-})
}} \approx F_{AB}\cdot
r_P
\eqno(9)   $$
Here the approximate equalities are exact if the tree contribution is
negligible, and will be violated where both contributions are
appreciable. In the latter case, each ratio is again equal to
an expression analogous to the right hand side of (3)
with appropriate different values for $T$ and $P$.

The SU(3)-breaking effect is different in decays to
final states containing $\eta$ and $\eta'$ because
the discrete transformation (Weyl reflection)
$d \leftrightarrow s$ which simply interchanges the d and s flavors
interchanges the $d \bar d$ and $s \bar s$ components in the
$\eta$ and $\eta'$ system, which we denote by $P_d$ and $P_s$. These decays are
of particular interest since recently reported high branching ratios\REF
{\CLEO}{B. Behrens, [CLEO] talk at $B$ Physics and CP Violation, Waikiki, HI
(March 1997)}${[{\CLEO}]}$ for
strange $B$ decays to $\eta'$ final states has led to suggestions for new types
of diagrams\REF{\ATSON}{D. Atwood and A. Soni, Phys. Lett. {\bf B405}
(1997) 150}${[{\ATSON}]}$.

We first note that the Cabibbo-favored tree diagram is expected to be dominant
in $\eta$ and $\eta'$ decays with charmonium and that in these decay modes they
are produced via $P_d$ in $B^o$ decay and via $P_s$ in $B_s$ decay. Thus these
decays immediately provide a measure of the $\eta - \eta'$ mixing. We first
obtain the SU(3) symmetry result
$$ {{A (B^o \rightarrow \psi^{(*)} P_d)
} \over {
A(B_s \rightarrow \psi^{(*)} P_s)
}} = r_{T(cds)}; ~ ~ ~ A (B^o \rightarrow \psi^{(*)} P_s) =
A(B_s \rightarrow \psi^{(*)} P_d) = 0
\eqno(10)   $$
This immediately gives the values of the strange and nonstrange components in
the $\eta$ and $\eta'$ and a condition which must be satisfied if the
$\eta - \eta'$ mixing is described by a $2 \times 2$ matrix.

$$ {{A (B^o \rightarrow \psi^{(*)} \eta' )
} \over {
A(B^o \rightarrow \psi^{(*)} \eta )
}} = {{\langle P_d \ket {\eta'}} \over {\langle P_d \ket {\eta}}}; ~ ~ ~
{{A (B_s \rightarrow \psi^{(*)} \eta' )
} \over {
A(B_s \rightarrow \psi^{(*)} \eta )
}} = {{\langle P_s \ket {\eta'}} \over {\langle P_s \ket {\eta}}}
\eqno(11a)   $$
$$ {{A (B^o \rightarrow \psi^{(*)} \eta' )
} \over {
A(B^o \rightarrow \psi^{(*)} \eta )
}} = - {{A(B_s \rightarrow \psi^{(*)} \eta )
} \over {A (B_s \rightarrow \psi^{(*)} \eta')
}} \eqno(11b)   $$
A failure of the relation (11b) would indicate a breakdown of the simple
mixing model.

We now investigate the decays into strange final states with
$\eta$ and $\eta'$.
The standard penguin diagram
predicts\REF{\PKETA}{Harry J. Lipkin, Phys. Lett. {\bf B254}
(1991) 247, and {\bf B283} (1992) 421}
${[{\PENGRHO,\PKETA}]}$
$$ \tilde \Gamma(B^\pm \rightarrow K^\pm \eta'): \tilde \Gamma(B^\pm
\rightarrow K^\pm \eta): \tilde \Gamma(B^\pm \rightarrow K^\pm \pi^o) =
3:0:1    \eqno(12a)                                          $$
$$ \tilde \Gamma(B^\pm \rightarrow K^{*\pm}(890) \eta'): \tilde \Gamma(B^\pm
\rightarrow K^{\pm*} \eta): \tilde \Gamma(B^\pm \rightarrow K^{*\pm}\pi^o) =
(1/3):(8/3):1    \eqno(12b)                                          $$
$$ {{\tilde \Gamma(B^\pm \rightarrow K^\pm \eta') + \tilde \Gamma(B^\pm
\rightarrow K^\pm \eta)}\over{\tilde \Gamma(B^\pm \rightarrow K^\pm \pi^o)}}
\leq 3    \eqno(12c)                                          $$
where $\tilde \Gamma$ denotes the theoretical partial width without phase space
corrections. We have assumed SU(3) symmetry with one of the standard mixings:
$$ \ket{\eta} = {1\over{\sqrt 3}}\cdot(\ket{P_u} + \ket{P_d} -\ket{P_s});
 ~ ~ ~ \ket{\eta'} = {1\over{\sqrt 6}}\cdot(\ket{P_u} + \ket{P_d} + 2\ket{P_s})
 \eqno (13)$$
and noted that the penguin diagram
creates the two states $K^\pm P_u$ and $K^\pm P_s$, with a relative phase
depending upon the orbital angular momentum $L$ of the final state.
$$A(B^\pm \rightarrow K^\pm P_s) = (-1)^L\cdot(1-\epsilon)\cdot
A(B^\pm \rightarrow K^\pm P_u)                       \eqno(14)  $$
where $\epsilon$ is a parameter describing SU(3) symmetry breaking
and $K^\pm$ can also denote any $K^*$ resonance.
The sum rule inequality (12c) holds generally for all mixing angles and for
all positive values of $\epsilon$.

The dramatic reversal of the $\eta'/\pi^o/\eta$ ratio in the final states with
$K^{*\pm}(890)$ occurs naturally in this penguin interference model and does
not occur in any other suggestion for enhancing the $\eta'$.
Present data indicate $K^{*\pm}(890) \eta'$ suppression.
Better data showing significant suppression will rule out most other
$\eta'$ enhancement mechanisms.

A violation of the inequality (12c) would require an additional
contribution. The Cabibbo favored charmed tree diagram
$A(B^\pm \rightarrow K^\pm P_c \rightarrow K^\pm \eta')$ can contribute
via hidden or intrinsic charm in the
$\eta'$ wave function and may contribute appreciably even though the
charm in the $\eta'$ is quite small.

We now estimate the effect of an additional contribution from the production
of the $\eta$ and $\eta'$ via an additional diagram which in the SU(3)
symmetry limit produces the states $\ket{P_u}$, $ \ket{P_d}$ and $\ket{P_s}$
with equal amplitudes.
$$ A(B^\pm \rightarrow K^\pm \eta) = \sqrt {2/ 3}\cdot \xi \cdot
A(B^\pm \rightarrow K^\pm \pi^o)                     \eqno(15a)  $$
$$
A(B^\pm \rightarrow K^\pm \eta') = \sqrt {1/ 3}\cdot (3 + 4 \xi) \cdot
A(B^\pm \rightarrow K^\pm \pi^o)                     \eqno(15b)  $$
$$
A(B^\pm \rightarrow K^{*\pm}(890)  \eta) = \sqrt {2/ 3}\cdot (2 -\xi) \cdot
A(B^\pm \rightarrow K^{*\pm} \pi^o)                     \eqno(15c) $$
$$
A(B^\pm \rightarrow K^{*\pm}(890)  \eta') = -\sqrt {1/ 3}\cdot (1 + 4 \xi) \cdot
A(B^\pm \rightarrow K^{*\pm} \pi^o)                     \eqno(15d)  $$
where $\xi$ defines the  extra contribution strength.
Consider for example the case $\xi=0.5$
$$ \tilde \Gamma(B^\pm \rightarrow K^\pm \eta'): \tilde \Gamma(B^\pm
\rightarrow K^\pm \eta): \tilde \Gamma(B^\pm \rightarrow K^\pm \pi^o) =
(25/3):(1/6):1    \eqno(16a)                                          $$
$$ \tilde \Gamma(B^\pm \rightarrow K^{*\pm}(890) \eta'): \tilde \Gamma(B^\pm
\rightarrow K^{\pm*} \eta): \tilde \Gamma(B^\pm \rightarrow K^{*\pm}\pi^o) =
3:(1.5):1    \eqno(16b)                                          $$
$$ {{\tilde \Gamma(B^\pm \rightarrow K^\pm \eta') + \tilde \Gamma(B^\pm
\rightarrow K^\pm \eta)}/{\tilde \Gamma(B^\pm \rightarrow K^\pm \pi^o)}}
\leq (17/2)    \eqno(16c)                                          $$
The inequality (16c) holds for all mixing angles and all $\epsilon \geq 0$.
Thus a comparatively small contribution interfering constructively with the
dominant penguin can give an appreciable enhancement.
With $\xi$ sufficiently large to give (25/3) for
($ \tilde \Gamma(B^\pm \rightarrow K^\pm \eta'):
\tilde \Gamma(B^\pm \rightarrow K^\pm \pi^o)$ and a 50:1 ratio favoring
$\eta'$ over $\eta$, the enhancement of $\eta'$ over $\eta$ is only a factor
of two for the $K^*$ final state.
The drastic difference between the $K$ and $K^*$ branching ratios still
persists if both the penguin and the extra contribution are present, in contrast
to the case where the extra contribution is dominant. Thus the $K^*$ data are
important for determining the exact mechanism for the $\eta'$ enhancement.

Discussions with M. Gronau and J. L. Rosner are gratefully acknowledged.

This work was  supported  in part by grant No.  I-0304-120-.07/93
from The  German-Israeli  Foundation for Scientific  Research and
Development and by the U.S. Department
of Energy, Division of High Energy Physics, Contract W-31-109-ENG-38.

\refout
\end